# The Role of Cloud of Things in Smart Cities

Sandip Roy, *Research Scholar, University of Kalyani, India*
Dr. Debabrata Sarddar, *Assistant Professor, University of Kalyani, India*

*Abstract*- The recent demographic trends indicate towards a rapidly increasing population growth and a significant portion of this increased population now prefer to live mostly in cities. In connection with this, it has become the responsibility of the government to ensure a quality standard of living in the cities and also make sure that these facilities trickle down to the next generation. A program named "Smart City Mission" has been started for this purpose. With an extremely diverse population, that is only second to China in the world in terms of size, the Indian government has engaged in serious thinking for a better city planning and providing numerous opportunities for the citizenry. It was, therefore, planned that the "Smart City Mission" program will be able to provide a highly responsive infrastructure, network security, a good living environment and the like. Internet of things (IoT) application in smart cities turns out to be the most challenging in this phase. The information available in the internet has made accessible to many devices through IoT and it also aware the citizen in many aspects. But with the increasing number of devices and information, it is now becoming increasingly difficult to depend on IoT to manage things in the internet space with a similar degree of ease. As a result, cloud-based technologies have given preferences over the existing one and IoT has been replaced by the newly introduced Cloud of Things (CoT) paradigm. This paper intends to connect different smart city applications for the betterment of city life with the Cloud of Things (CoT). Our proposed smart city architecture is based on Cloud of Things, and the focus is also given to identify the existing as well as the forthcoming challenges for the concerned program of the government. By identifying the difficulties it is expected that the project will be materialized with a great success.

I. INTRODUCTION

The Modern city life cannot be imagined without internet (namely IoT). And cloud computing provides an efficient way of maintaining resources in the internet [1, 2]. We start with the origin of IoT in the modern world. Kevin Ashton, in 1999 first thought of IoT [3]. At that time, the role of IoT was limited to only building connection between interacting objects in the server over the internet. In the last few years, as a result of technological advancement in sensor deployment mostly for big data center and increasing use of cloud technology, the sensor deployment in the internet has become crucial to think over [4]. There is no doubt that the use of internet enabled devices is increasing day by day. It is expected that 50 to 100 billion devices will be connected with the internet within 2020 as forecasted by the European commission [5, 6]. In fact in the year 2008 these numbers already exceeds world's population half of which are now living in different cities all over the world [6]. In Fig. 1 we have made a comparative analysis between the numbers of internet connecting devices to that of world population from the year 2008. It is clear that the former exceeded the latter already in the year 2008 and later on also there is no exception in this trend. This has a direct effect of making the city life more attractive as compared to rural one. It may be, therefore, inferred steps towards smart city project by the government and different private organizations are quite rational [7]. Therefore, internet connected devices are attracting the attention of many researchers.

The projects under smart city mission are not uniform all over the world. With the passes of time and with the change in the living standard the various needs of people are also changing [8, 9]. Development in different aspects such as in the fields of infrastructure, agriculture, electricity supply, energy storage, railways, waste water management, management of traffic control and safety measure for men and women vary from country to country. It becomes evident that the concept of smart city is not universal [10]. Being a developing country, India has taken an initiative of creating a mission of developing hundred cities to smart cities [11, 12].

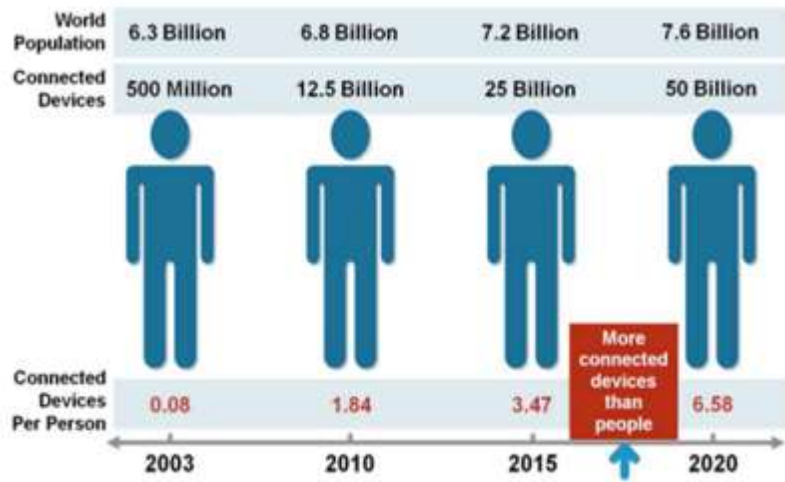

Figure 1. Year wise measures of connected devices with the world population

Fig. 2 represents the allotted budget for investing in different areas undertaken in the project sanctioned by the finance minister of India and the underlying projects are presented in Fig. 3.

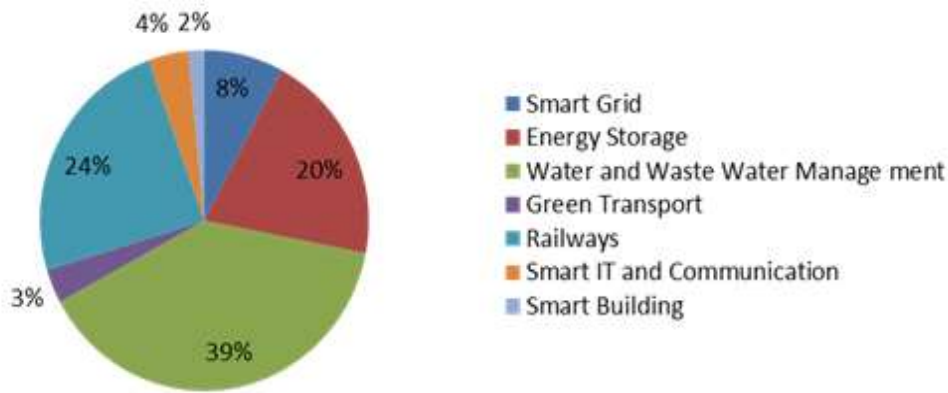

Figure 2. Investment plan in different sectors by Government of India for Smart Cities

For accomplishing the smart city mission, IoT has been used in the initial stages to manage data as it works through Wireless Sensor Network (WSN) for collection of data through sensors or even aggregated from humans. The data can be visualized in the web or even exchanged if necessary. But with the passes of time the use of internet enabled devices has increased rapidly entailing a hung amount of raw or unstructured data which cannot be efficiently handled by the IoT alone. The cloud technology here gives a way out. So, an integration between the cloud computing and IoT is required for handling this unstructured data efficiently [13, 14]. This integrated service is termed as Cloud of Things (CoT). The role of cloud in CoT architecture is analogous to a central control unit that processes the raw data in a systematic manner [15]. Here the sensor nodes first sensed the data, followed by data processing in the cloud and as shown in the physical world [15, 16]. In this paper, we discuss the design of CoT to reflect upon the advantages of clod technology and its ability to work beyond the limitations of IoT.

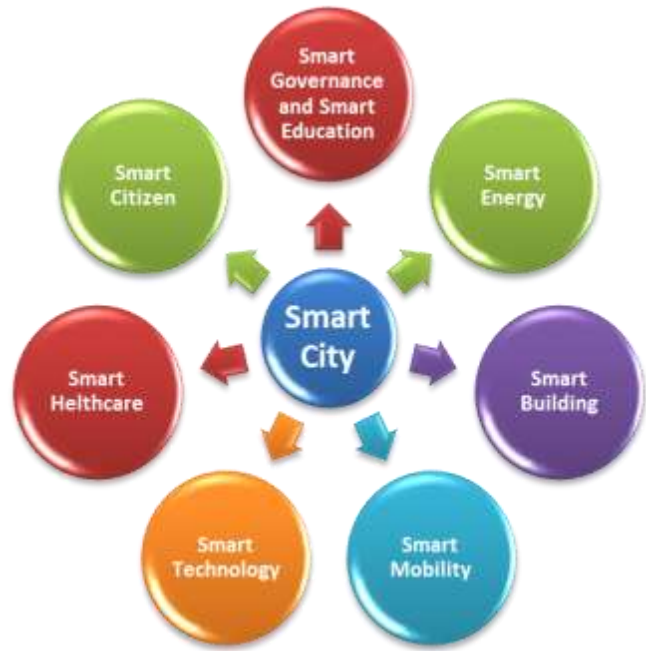

Figure 3. Projects diversity in Smart City

In the rest of the paper we focus on the following aspects: in section 1 we analyze the current trends towards the smarter world focusing on the service model of cloud termed as Everything as a Service (XaaS) and also tried to depict a picture of expected market growth of global cloud computing and XaaS service model in the year 2016. The investment planning for smart cities are explained is section 2. Section 3 deals with detail architecture of the smart city based on Cloud of Things. In section 4 we review different application of CoT in smart cities. A brief overview of IBM smart cities is presented in section 5. Finally the limitations and upcoming challenges for the concerned project in a developing country like India is viewed in section 6 and the concluding remark has made in the last section, section 7.

II. STEPS TOWARDS SMART CITY

Today the government, information and communication technology as well as some private sectors are taking interest in the smart city mission. In addition to the basic needs a clean and sustainable environment and application of smarter technology have now also become the priorities of modern urbanized people [17, 18]. To meet these requirements technologists first prefer different IoT based applications [19]. The attention has given to different aspect of smart city solutions which include development of public transportation, reduction of traffic congestion, promote agriculture, waste water management, efficient energy grid and methodology for keeping safe of the citizen life [20]. But IoT alone cannot fully explore all these areas. So technologists require a collaboration of IoT with cloud of Things [21]. Dr. Ramnath Cheliappa in 1997 make us familiar with cloud computing and soon its successful application in different software industry make it popular [22]. Salesforce.com (1999), Amazon Web Service (2002), Elastic Compute Cloud and Simple Storage Service (2006) are some of the few services that capture the market in a large scale [23, 24].

In cloud computing resources such as software and hardware are deployed to a physical data center. Different users can access those resources as services irrespective of their geographical location on the basis of internet connection. There is an upward rising trend in the use of cloud technology from the very beginning especially since 2012 and today cloud captured 95% of the market share. Of which the major share being that of hybrid cloud (71%) whereas the share of private and public cloud being 6% and 18% respectively [25]. The following Fig. 4 presents the history of cloud computing in an efficient manner:

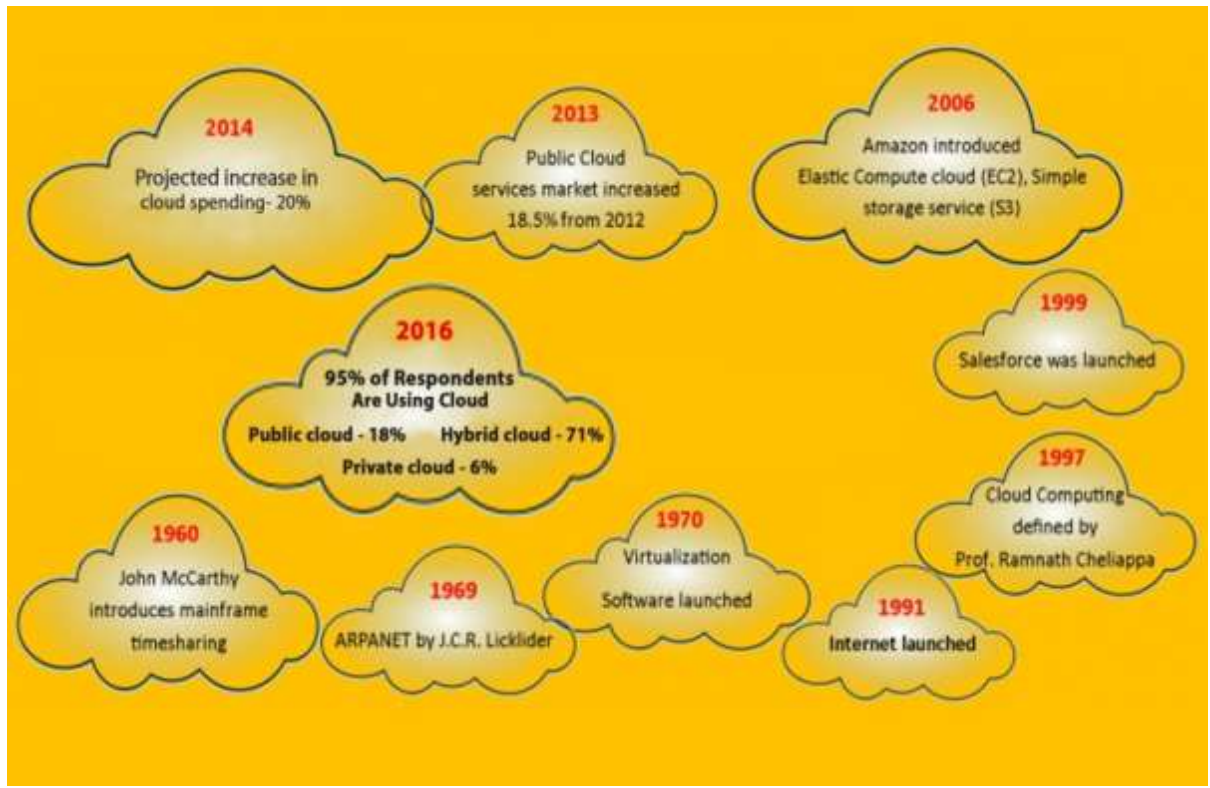

Figure 4. The history of Cloud Computing

Platform as a service (PaaS), Infrastructure as a service (IaaS), Software as a service (SaaS) are the primary service models of cloud computing [26]. With change in the pattern of demand several other services are added to cloud and make accessible to the users by the service provider on the payment basis. Consumers only pay for the amount of services that they use. All these services are brought under a new paradigm named everything as a service (XaaS) [6]. The service models in the cloud are shown in Fig. 5:

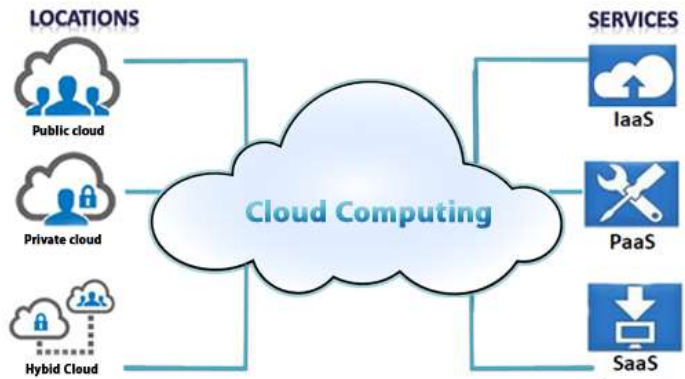

Figure 5. Service models of Cloud Computing

Statistical analysis suggests global cloud computing and XaaS markets will show a sharp rise from $37.8 billion in 2010 to $121.1 billion in 2015. Along with this the global smart city market is expected to exceed $1 trillion by 2016 [6]. This growth analysis is represented in Fig. 6 from the year 2010 to 2016. The private organizations involved in the smart city mission are IBM Smart Planet and Smart Cities, Dubai Smart City, European Smart City and Oracle Government and many more. The governments as well as these private companies have already started working towards the Smart city Project [18, 27 – 29].

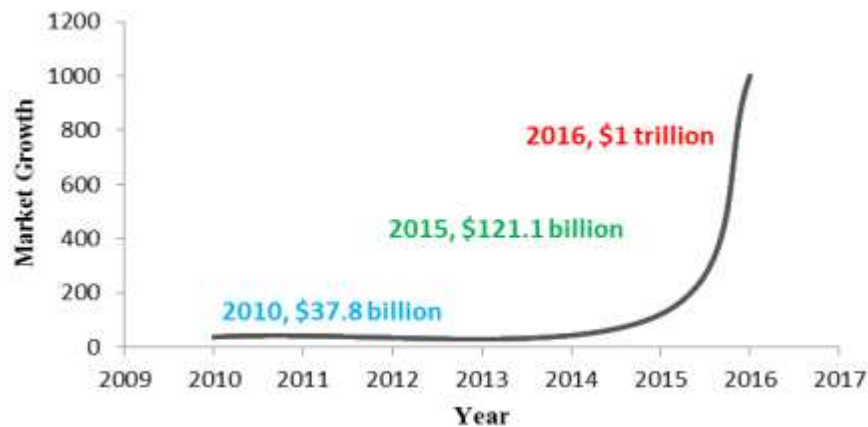

Figure 6. Expected market growth rate of XaaS market

A. *Plan for Smart Cities in India*

As a result of improvement in city life immigration to cities has taken place in a large number. In India it has been predicted that nearly 25-30 people will migrate every minute from their rural or sub urban residents to major Indian cities in search of a better quality of living [30]. With this rate of migration the number of people living different Indian cities is expected to reach 843 million [31]. The urbanization of Indian citizens is shown in Fig. 7 below:

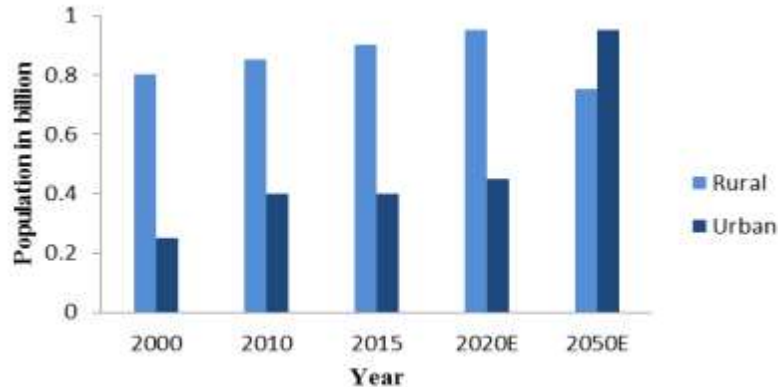

Figure 7. India space of urbanization

In the fiscal year 2014-15 the Government of India has allocated ₹980 billion (US$15 billion) for developing 100 smart cities in India [32]. The decision to build up modern satellite towns has also been undertaken in the project. The expected amount require for the project is 1.2 trillion over the next twenty years [11]. Materialization of the smart city plan demands a considerable number of man power which results in employment generation in both public and private sectors. Approximately 10-15% employment is expected to rise due to Smart City Plan. Investment decision for different areas of smart city has already been initiated. These areas include Smart Grid, Smart Environment, Waste water management, Smart Transportation, Smart Home and Smart IT and communication.

### III. THE DETAILED INVESTMENT PLANS FOR SMART CITIES IN INDIA

Starting with Smart Grid, a plan for eight pilot smart grid projects has decided to be undertaken for which an estimated $10 million is projected to invest. Along with this a power set up so that all houses in the smart cities should have at least electricity connection for 8 hours/day is taken into consideration [33, 34]. Presently India is lacking the infrastructure required for this connection and the deadline to meet these requirements is decided to be 2017. By 2017 India needs eighty eight thousand MW of power generation capacity and also minimum 250-300 GW new power generation station within next thirteen years. No doubt this demands for a huge amount of investment and the Power Grid Corporation of India Ltd has planned to invest $26 billion for the next five years [35]. Besides power generation installation of 130 million smart meters by 2021 is also a part of power grid mission [36].

In connection with rapid environment pollution any mission for better living cannot be completed without considering for a cleaner environment plan. Smart City Mission is also no exception. Focus has been given to achieve a clean and pollution free environment. The first step in this direction is a shift towards new and renewable resources for the required power generation (eighty eight thousand MW). An approximate $50 billion is to be invested foe this by the Ministry of Water Resources within next year [11]. For improving the means of transportation attention is given towards the production of electric and hybrid vehicle. The government sanctioned $4.13 billion for the smart transport project and has already settled six million vehicles by 2020 [11]. To recover infrastructure cost electric vehicle charging stations are to be set up in all urban areas along all states and National

Highways by 2027. As a means of quick transportation $20 billion is projected to invest on metro rail projects and $10.3 out of government revenue has to be allotted for high speed railway from Mumbai Ahmedabad [11].

For smart IT solution and communication Digital India mission is taken. Whenever IT industry is considered the role of cloud computing in recent years cannot be ignored. So, spreading up the applications of cloud technology in different industries is a part of IT project. Expectedly cloud will be evolved in $4.5 billion markets in India this year and also for increasing the use of internet nearly 175 million users will be connected with internet through broadband connection by 2017 [11]. The people in the smart cities will live in smart building in which they can save near about 30% energy use, 10% water use and roughly 10 to 30% in the maintenance cost [33]. For setting up smart building the investment amount sanctioned by the Intelligent Smart Building Management is $621 million which will rise up to $1891 million by 2016 [11]. The investment revenue for the smart city project is shown in the following Fig. 8:

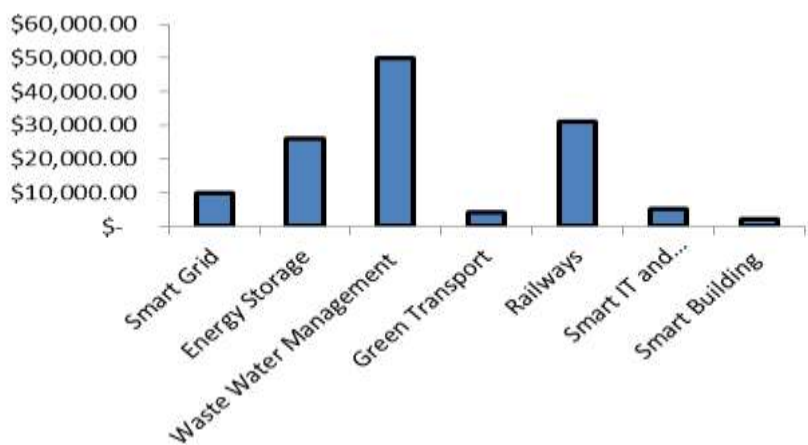

Figure 8. Detailed investment revenue for Smart cities in India

IV. AN ARCHITECTURE OF SMART CITY BASED ON CLOUD OF THINGS

In this section we will discuss the basic layered architecture of smart city (Fig. 9) and the layered architecture of cloud of things (Fig. 10). Smart city project is a good initiative for giving the citizen a better life with current trend of urbanization, concentration of population in urban areas and to overcome the obstacles in infrastructures [7]. Obviously the foundation of smart cities lies in smart Information and Communication Technology (ICT).

The architecture of smart city consists of four layers namely Infrastructure, E-governance, Services and Stake holders [7]. For each layer some related components have discussed below.

For infrastructure layer sensors/actuators, high end data centers, different wired and wireless network connectivity between different smart devices. Coming to E-governance layer, it comprises of different public departments with their respective activities. These departments have the responsibilities for efficient planning and management of the cities. So, maintaining connections between departments are necessary for successful implementation of smart city plan. With smart information and communication technology the smart city project has now become more efficient, technologically sound and most importantly citizen friendly. The Service layer offered different public services associated with the needs of the citizens. These services include waste water management, smart grid, green transport, smart IT and communications. All the services are provided to the citizens as well as to the government officials who are considered as stake holder in the smart city architecture in a convenient manner.

Each government department is working independently and share limited information to each other. But with ICT intervention the information are not confined to the respective department only but there is an open platform or open data model where the other interested parties can access the information if required.

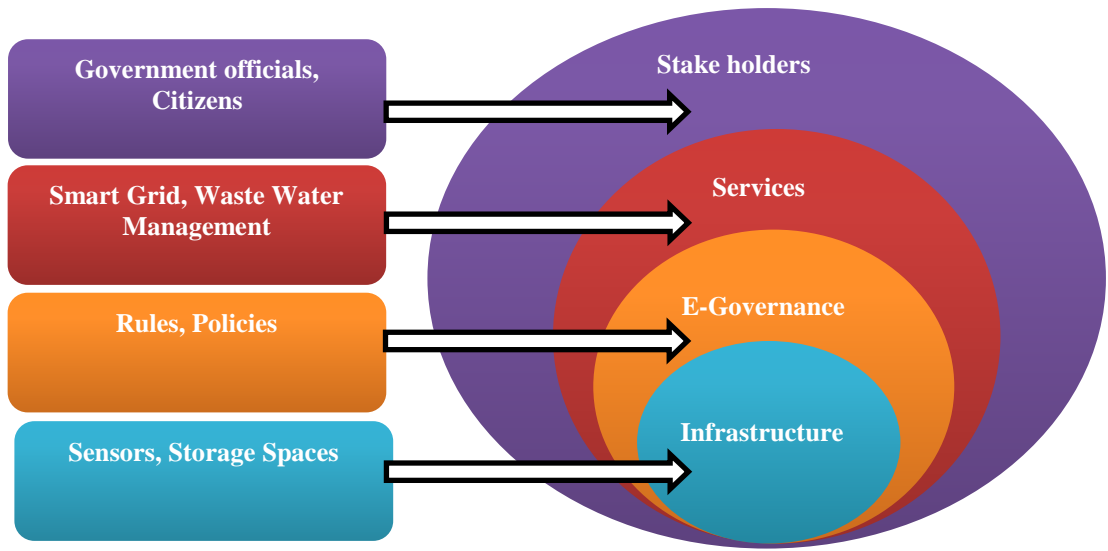

Figure 9. Layered architecture of Smart City

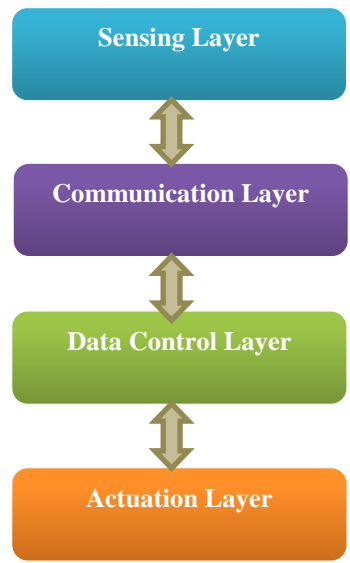

Figure 10. Layered architecture of Cloud of Things

Again for cloud of things architecture comprises four layers [15]. The first 2 layers i.e. the Sensing and Communication layer perform the same task as infrastructure layer does in smart city architecture [15]. Start with data control layer which is an important one in this architecture. Data collected in the first two layers from different sources are in the form of unorganized information which needs to be processed so that it can be used at times of need. The information is then sent to the data control layer via wired and wireless network. All the data are stored and processed here. The data control layer helps to build up a connection between the service providers and the vend users to access different services on their devices. The working in this layer is similar to those discussed in the service layer in the previous smart city layers. The final stage in the architecture is actuation layer which represent the processed data collected from the control layer either manually or in automatic. If users choose manual representation then they have to intervene and do the work accordingly. But in case of automatic set up human intervention is required rather the actuators works in accordance with own given instructions.

To sum up the architecture section, we see that the smart city architecture based on Internet of Things is not compatible with present requirement of storage space and processing power at a large scale. But this can be complemented with incorporating cloud of things to provide complete services to the targeted population. The data control layer discussed above is able in filling the gaps left by IoT (Fig. 10, Fig. 11). There is no problem of data storage, sharing, fusion and application of suitable algorithm results in appropriate and sophisticated actions. So the benefit of our detailed proposed architecture (Fig. 11) is not only limited to urban citizens but suits perfectly well to the modern technological evolution.

## V. Cloud of Things based on Smart Cities

In this section we consider some direct applications of cloud of things. In the field of machine to machine communication (M2M) the evolution started with the introduction of internet of things has been increased many times with the applications of cloud of things [15, 37]. Below are the different areas of application of proposed technology.

### A. Smart Energy and Smart Grid

The use of IoT for energy distribution and consumption is limited in the sense that it can process data only at a small scale. Here comes the role of cloud which facilitates computing data in a large scale. Computing resources in cloud server provides a secure and reliable networking system with a self-dependent mechanism. A huge amount of data and information collected from different sources are handled by the cloud computing technique efficiently [38].

### B. Smart Environment Monitoring

For fulfilling the objective of a cleaner environment monitoring environment is crucial. Cloud is applied in areas where direct monitoring is difficult. Monitoring problem is faced particularly in case of non-point source polluting areas like gas concentration in air inside mining, laboratories and vehicles. Besides the collection of continuous and longtime sensor data require sufficient storage space and efficient management and processing technique which can be provided by CoT [19].

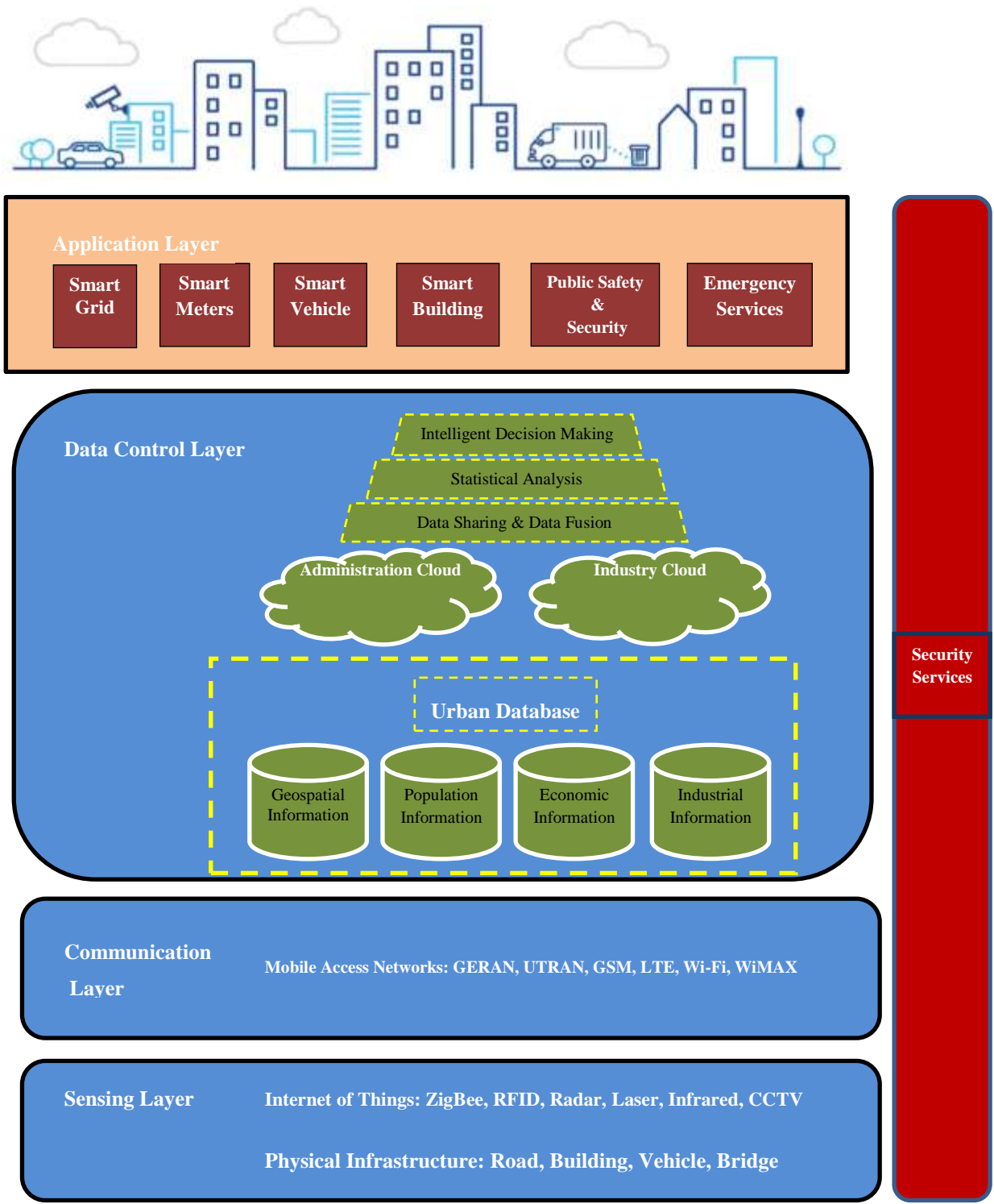

Figure 11. Smart City Architecture based on Cloud of Things

*C. Smart Logistics*

Adequate food requirement is one of the basic needs of human. With increasing population the demand for food is also increasing day by day. To meet this demand in addition to raising the food production, prevention of food loss is also needed. Smart technologies here come out to be very effective. Various smart logistics like pack houses and refrigerated vehicles prevent food loss to a great extent and optimized its use. In addition, in recent market the introduction of supply chain is very useful. It not only lead to efficient marketing of food produced but also make it possible for the farmers to reach to the end users [39]. To fulfill the objective of meeting food demand India needs to increase the number of well-equipped pack houses from 250 in present to 70000 in the near future [40].

*D. Waste Management*

For a cleaner city waste management is very important. Proper and systematic waste management involve a chain of process. For example, firstly the waste should be collected on a regular basis, then the collected waste needs to be transported to a proper place where the waste can either be recycled or degraded according to their types, finally the processing, managing and monitoring the waste material come under this chain. This traditional process require considerable amount of money, time and labor. So, optimizing waste management leads to reduction in expenditure that can be spent on other development purpose [6]. To optimize waste management the responsible authorities require strategic movement. The sensors data can help a lot in formulating garbage collection strategy. If this happen then it will be possible to save fuel cost of garbage trucks. These data can also be analyzed by different recycling companies for predicting and tracking the garbage coming to their plans so that they can make the processing plan accordingly. Automatic monitoring will contribute to further cost saving by reducing the complexities of manual monitoring and also leads to time saving. All these based on cloud of things technology. Therefore application of CoT in waste management saves revenue and this revenue can be allotted to other areas of smart cities.

VI. A SCENARIO FOR UNDERSTANDING IBM SMART CITIES

Fig. 12 represents the key areas which IBM decided to focus for their smart city project. Here we considered five key facts that are directly related with smart city application. These key areas are as follows: 1) Smarter Water, 2) Smarter Public Safety, 3) Smarter Traffic, 4) Smarter Buildings, and 5) Smarter Energy [27].

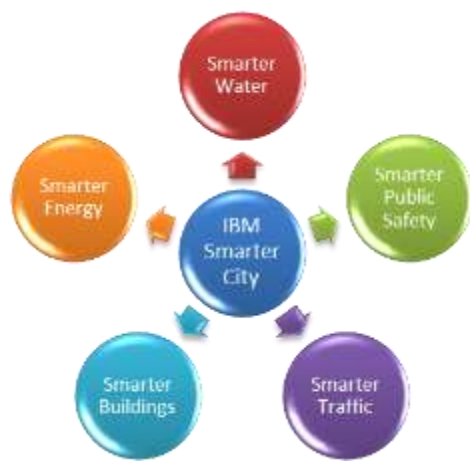

Figure 12. IBM Smarter City

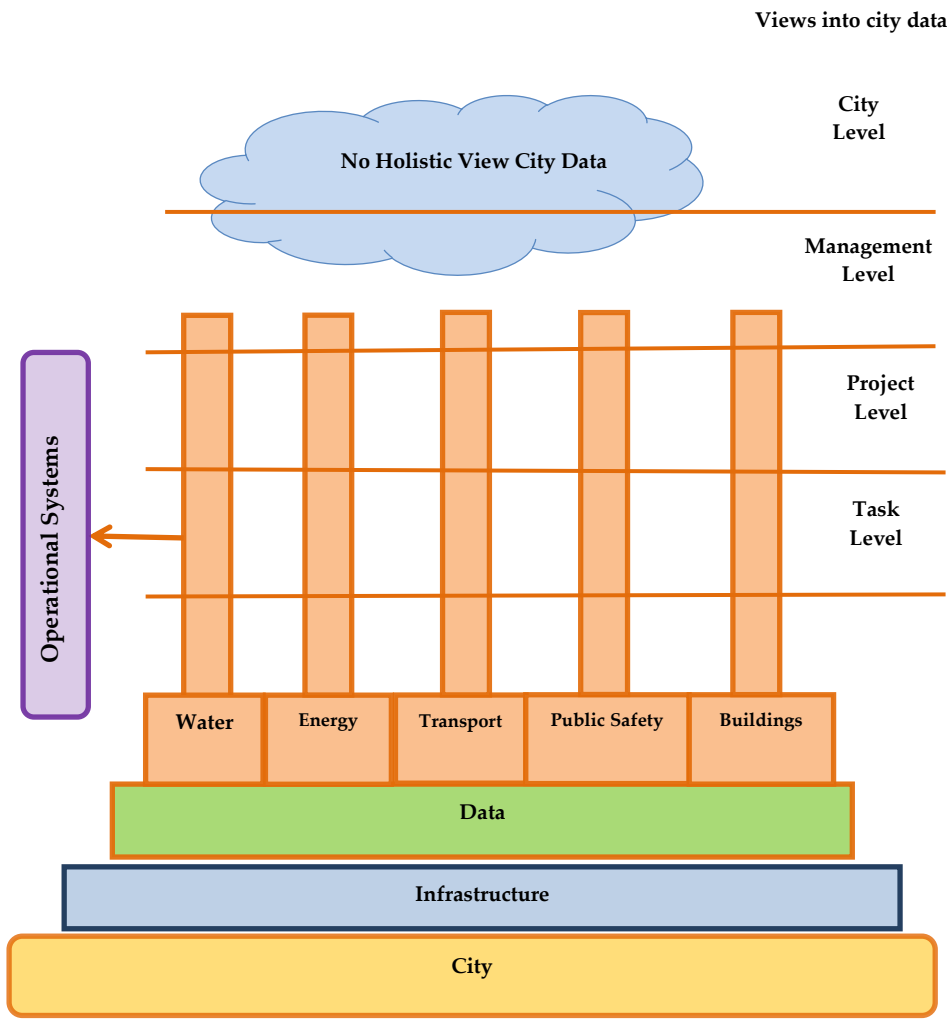

Figure 13. Traditional domain of a City

In the two figures, Fig. 13 and Fig. 14 describe the traditional and holistic view of a smart city [41]. In a traditional city in Fig. 13 there are different organizations and departments with their respective duties and provide definite services to the people living in the concerned city. Each department deals only with its own responsibilities, the problems faced on that particular domain and there are no strong connections between different departments as each regulate its work independently. Also any interested third party can only have limited information from any department. So, there is lack of access to clear information for which traditional infrastructure and services face several obstacles in moving towards a smarter city.

In the following Fig. 14 we have a holistic view of smarter cities. It gives a complete understanding regarding the problem faced before taking any decision. The city model is designed in such a manner that individual domains are optimized in real time. In addition it can monitor, interconnect or control the domains either separately or taking all the domains altogether unlike the traditional view of the city where each department works independently. All the information coming from different domains is collected in a single center named cross domain operation center. Also there is domain specific operation centers where data and information from individual domain is stored and can be used for betterment of any planned (such as Olympic Games, World Cub, and Indian Premier League) or unplanned events (such as floods, earthquakes, and tsunamis).

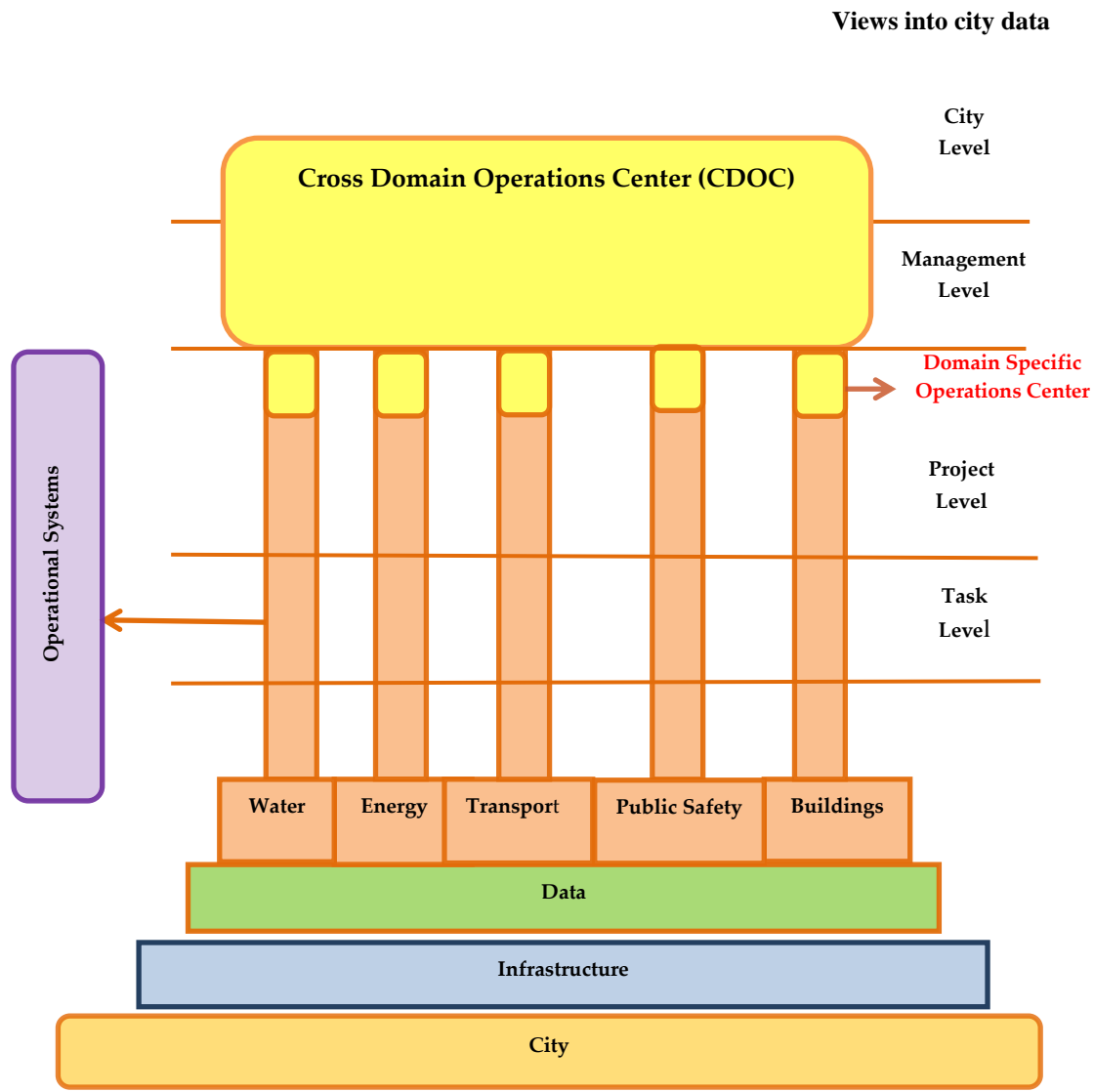

Figure 14. Holistic views of Smarter Cities

VII. LIMITATIONS AND CHALLENGES FOR DEVELOPING SMART CITIES IN INDIA

Even though the plan for smart cities seems good in paper there are some problems in practical execution of the plan in a developing country like India. It is expected that the smart city mission leads to economic growth, employment opportunities in different sectors along with development goals like increasing the education level, skills and income level of the citizens. But a number of cities are now suffer from various problems such as lack of water supply, sable electric power ,poor condition of educational institution and the like. The Indian government has undertaken a plan for developing hundred smart cities but it does not have enough revenue to materialize the plan. So, challenges coming from both technical and privacy view point. Some of these are discussed below:

*A. Unable to manage unstructured data*

In smart cities there are increasing numbers of devices connected through IoT. Now data generated from different applications are mostly unstructured. Proper use of these data requires managing and storing them efficiently. Again efficient management needs suitable infrastructure which is presently lacking [42].

*B. Inefficient storage space*

Another related problem of unstructured data is lack of storing capacity of numerous data. The sensors data are of no use if it cannot be stored efficiently. In fact the huge amount of data without suitable storage space increases pressure on the existing system and creates further problem in processing the data [7].

*C. Insufficient processing capability*

IoT is incapable of processing and handling huge amount of data. This lack of processing capacity causes the users of the smart devices to suffer from various problems [7].

*D. Privacy and Security issue for data handling*

Data coming from any devices or users should be protected from third party intervention. For this proper authentication is required. Rules and laws related to the security issue needs to be published to the users to prevent unauthorized access. The incident of cyber-crime is a very common phenomenon today. Any information should be shared only after granting permission from authentic users [42].

## VII. CONCLUSION

The survey paper starts with analyzing the rationale for smart city project in recent days. Step by step it considers different aspect for a successful development of smart city mission. It represents the estimated investment plan for smart city in detail. A 'four layered architecture' of smart city based on Cloud technology is presented and the applicability of cloud for improving the traditional architecture is also viewed in brief. A comparative analysis of traditional and holistic view of smart city is made with diagrammatic representation. In the holistic viewpoint the usability of cross domain data operation centers and domain specific operation centers are considered for smarter cities architecture [41]. At the end of the paper the limitations and challenges in developing smart cities are summarized. The focus of the survey paper is on detailed architecture of smart cities. But the functional representation of each layer, the mechanism of communication or connection between the layers for sharing data and the related discussions are beyond the scope of present study.

ACKNOWLEDGMENT

All Authors gratefully acknowledge to Computer Science & Engineering Department of University of Kalyani, Kalyani and Sabita Devi Education Trust - Brainware Group of Institutions, Kolkata, India, for providing lab and related facilities for do the research.

ABBREVIATIONS

The following abbreviations are used in this manuscript:

| | |
|---|---|
| CoT | Cloud of Things |
| XaaS | Everything as a Service |
| ICT | Information and Communication Technology |
| IaaS | Infrastructure as a Service |
| IoT | Internet of Things |
| PaaS | Platform as a Service |
| SaaS | Software as a Service |
| WSN | Wireless Sensor Network |